\documentclass{ws-procs975x65}
\usepackage{graphicx}

\def\beq{\begin{equation}}
\def\eeq{\end{equation}}

\begin{document}

\title{ General-relativistic rotation laws in rotating fluid bodies, new weak-field effects  and the post-newtonian expansion}
\author{Patryk Mach, Edward Malec$^*$ and Micha\l~ Pir\'og}

\address{ M. Smoluchowski Institute of Physics, Jagiellonian University\\
Cracow, Poland\\
$^*$E-mail: Edward.Malec@uj.edu.pl\\
www.uj.edu.pl}

\begin{abstract}
Recent general-relativistic extensions of Newtonian rotation laws for self-gravitating stationary  fluids
allow one to rederive, in the  first post-Newtonian approximation, the well known geometric dragging of frames, and two new   weak-field effects within rotating tori. These are the recently discovered   anti-dragging   and a new effect that measures the deviation from the Keplerian motion and/or the contribution of the 
fluids selfgravity.   They can be applied to the study of    the existence of  the (post-)Newtonian limits of   solutions and  in investigations of    inequalities relating parameters  of rotating  black holes.\end{abstract}


\bodymatter


\section{Introduction}

Stationary Newtonian hydrodynamic  configurations --- axially symmetric and selfgravitating ---  that  rotate around a fixed axis, possess two pecularities. This is a free-boundary problem ---  its domain  has to be found  simultaneously  with the solution itself.  In addition, the problem is undetermined  ---  traditionally one fixes it by  (almost) freely prescribing  the so-called rotation curve, the angular velocities $\Omega_0$ of the particles of a fluid. There exists an integrability condition  ---  that  $\Omega_0$ and the   angular momentum per unit mass  $j$ can be functions of a single variable $r$, where $r$ is the distance from the rotation axis.  The  additional ``stability'' restriction $\frac{dj}{dr}\ge 0$  of Ref.\ \refcite{Tassoul}    constrains   freedom to choose the angular velocity  rather weakly. 

General-relativistic counterparts of these systems are also undetermined, but the way to   specify  the system is through prescribing the function $j(\Omega )$ ---  the angular momentum as a function of the angular velocity. This condition can be rephrased in principle in terms of geometric coordinates, but only through solving a nonlinear equation that  defines the  linear velocity. This vague remark  shall be explained later; we want only to say that it is    nontrivial  to find an admissible form of $j(\Omega )$.  The only  known  rotation law in general-relativistic hydrodynamics had been for a long time  that propoposed by Bardeen and Wagoner \cite{Bardeen_Wagoner}, with $j$ being a linear function of the angular velocity.   Galeazzi, Yoshida and Eriguchi \cite{GYE} have found  recently a nonlinear angular velocity profile. None of these tends in the Newtonian limit into    the Newtonian monomial rotation curves $\Omega_0=w/r^\lambda $. In this talk we shall introduce     general-relativistic rotation curves $j=j(\Omega )$  that in the nonrelativistic limit exactly coincide with   $\Omega_0=w/r^\lambda $ ($0 \le \lambda  \le 2$; $\lambda $ and $ w$ are constants).  They comprise, in particular,    the general-relativistic Keplerian rotation law that possesses the first post-Newtonian limit (1PN). The well known  Keplerian massless disk of dust satisfies our rotation law exactly in the Schwarzschild spacetime.

We shall consider   fluids with the  polytropic equation of state  $p(\rho ,S) = K(S) \rho^\gamma$,
where $S$ is the specific entropy of fluid. Then one has specific enthalpy  $h(\rho ,S) = K(S) \frac{\gamma}{\gamma-1}\rho^{\gamma-1}$. The entropy is assumed to be constant.

\section{Rotation laws in Newtonian hydrodynamics}
 
 The  Euler  equations read  
\begin{equation}  -\Omega_0^2 r=-\partial_r  (U_0 +h), ~~~ 0=-\partial_z  (U_0 +h);
\label{EN}
\end{equation}

 One easily notices    that   this system of equations, supplemented by the Poisson equation $\Delta U_0=4\pi G\rho $, is closed  only if $\Omega_0$ is known. Differentiation of the two equations in (\ref{EN}) 
with respect to $z$ and $r$, respectively, and subtraction of the results yields $\partial_z\Omega_0=0$.
The angular velocity depends only on  $r$, the distance from a rotation axis: $\Omega_0= \Omega_0(r) $.
This constitutes  the integrability condition.

\section{Equations of general-relativistic hydrodynamics}

We assume the axially symmetric  \emph{stationary} metric 
\begin{align}
\label{metric}
d s^2 &=  -  e^{\frac{2\nu }{c^2} }(d x^0)^2
+r^2   e^{\frac{2\beta }{c^2} } \left( d \phi  -  \frac{\omega }{c^3} d x^0\right)^2
+e^{\frac{2 \alpha }{c^2} }   \left( dr^2 +   dz^2\right)    .
\end{align}
Here $x^0 =ct$ is the rescaled  time coordinate ($c$ is the speed of light in vacuum), and $r$, $z$, $\phi$ are cylindrical  coordinates. The metric potentials $\nu$, $\beta$, $\omega$ and $\alpha$ depend on $r$ and $z$ only.   
 
The Einstein equations, with the signature of the metric $(-,+,+,+)$, read
$
R_{\mu \nu} -g_{\mu \nu }\frac{R}{2} = 8\pi \frac{G}{c^4}T_{\mu \nu },
$
where $T_{\mu \nu }$ is the stress-momentum tensor.  
 We   take the stress-momentum tensor  
$
T^{\alpha\beta} = \rho (c^2+h)u^\alpha u^\beta + p g^{\alpha\beta},
$
where $\rho$ is the baryonic rest-mass density, $h$ is the  specific enthalpy,
and $p$ is the  pressure.  
The 4-velocity  {$u^\alpha  $}  is normalized, $g_{\alpha\beta}u^\alpha u^\beta=-1$.
 The coordinate (angular)  velocity reads ${\vec v}= \Omega \partial_\phi $, where $\Omega = u^\phi /u^t$.
  
  Define the   linear velocity with respect to the locally nonrotating system of reference
 \begin{equation} 
 V=r \left( \Omega -\frac{\omega }{c^2}\right) e^{\left( \beta - \nu \right)/c^2}.
 \label{V}
 \end{equation}
The potentials $\alpha$, $\beta$, $\nu$, and $\omega$ satisfy  equations that have been found by Komatsu, Eriguchi and Hachisu \cite{komatsu}.
Here we recall a version similar to that used in Ref. \refcite{nishida_eriguchi}. The relevant equations read
\begin{eqnarray*}
\Delta \nu  &=&  4 \pi \frac{G}{c^2} e^{2 \alpha/c^2} \left[ \rho (c^2 + h) \frac{1 + V^2/c^2}{1 - V^2/c^2} + 2 p\right]  + \frac{1}{2 c^4} r^2 e^{2(\beta - \nu)/c^2} \nabla \omega \cdot \nabla \omega \nonumber \\
& & - \frac{1}{c^2} \nabla (\beta + \nu) \cdot \nabla \nu
 \nonumber\\
\left( \Delta + \frac{2}{r} \partial_r \right) \omega  & = & - 16 \pi \frac{G}{c^2} e^{2 \alpha/c^2} \rho (c^2 + h) \frac{\Omega - \omega/c^2}{1 - V^2/c^2}   + \frac{1}{c^2} \nabla (\nu - 3 \beta) \cdot \nabla \omega,
\end{eqnarray*}
where $\nabla$ denotes the ``flat'' gradient operator.   
They constitute an overdetermined, but consistent,  set of equations.  We omit the remaining Einstein equations, since they yield corrections of higher orders. The general-relativistic Euler equations are integrable assuming  that    the angular momentum per unit mass,
 \begin{equation}
j  =  u_\phi u^t= \frac{V^2}{\left( \Omega -\frac{\omega }{c^2}\right) \left( 1-\frac{V^2}{c^2}\right) },
\label{j}
\end{equation}
depends only on the angular velocity $\Omega $; $ j\equiv j(\Omega )$. For a given $j(\Omega )$, one can insert Eq. (\ref{V}) into Eq. (\ref{j}) and find explicit dependence of the angular 
velocity on spatial coordinates. It is clear, that this depends crucially on the form of $j(\Omega )$.

Assuming  $ j\equiv j(\Omega )$ one gets  a general-relativistic integro-algebraic Bernoulli equation, that    embodies  the hydrodynamic information carried by the continuity equations  $\nabla_\mu T^{\mu \nu }=0$ and the baryonic mass conservation
$\nabla_\mu \left( \rho u^\mu \right) =0$.  It is given by  
  $\ln \left( 1+\frac{h}{c^2}\right) +\frac{\nu }{c^2} +\frac{1}{2}\ln \left( 1-\frac{V^2}{c^2}\right) +\frac{1}{c^2}\int d\Omega j(\Omega ) =C.$

\section{Rotation laws in general-relativistic hydrodynamics}

The general-relativistic rotation law employed in the literature Refs.\ \refcite{Bardeen_Wagoner}, \refcite{komatsu}--\refcite{nishida1},      has the form
$
 j(\Omega ) = A^2 ( \Omega_c -\Omega )
 $,
 where  $A$  and $\Omega_c$ are  parameters. In the Newtonian limit and large $A$ one arrives at the rigid rotation, $\Omega = \Omega_c$, while for small $A$ one gets the constant angular momentum per unit mass.  A three-parameter expression for $j$ is proposed in Ref.\ \refcite{GYE}. None of these rotation curves give in the Newtonian limit the monomial 
 ones.
   
Two of us have   found recently  a new family of rotation laws \cite{MM},
  \begin{equation} 
\label{momentum}
j(\Omega ) \equiv \frac{w^{1-\delta }  \Omega^\delta }{1-   \frac{ \kappa  }{   c^2 }  w^{1- \delta }\Omega^{1+\delta } +\frac{\Psi }{   c^2}  },
  \end{equation}
 where $w$, $\delta ,  \kappa =(1 - 3\delta )/(1 + \delta ) +  \mathcal{O}(c^{-2}) $ and $\Psi $ are   parameters. 
 
It is notable that a  massless disk made of dust   rotating in Keplerian motion   is an exact solution  in the Schwarzschild spacetime.
 The  general-relativistic Bernoulli equation   acquires  a simple algebraic  form \cite{MM}, if $\delta \neq -1$:
\begin{eqnarray}
  \left( 1 + \frac{h}{c^2}\right)  e^{\nu /c^2}   \sqrt{1-\frac{V^2}{c^2}} \times  
\left( 1 - \frac{ \kappa  }{   c^2 }  w^{1- \delta }\Omega^{1+\delta } +\frac{\Psi }{   c^2} \right)^{\frac{-1}{\left( 1+\delta \right) \kappa }}   =  C.
\label{algebraic_Bernoulli}
\end{eqnarray}

  The seemingly singular case $\delta =-1$, that corresponds to the constant  linear velocity, is also described by  the present formalism;  one should take a limit $\delta \rightarrow -1$ in Eq. (\ref{momentum}) and some of forthcoming formulae (Ref.  \refcite{KMM}).

 Rotation curves $\Omega \left( r, z \right) $ ought to  be recovered from  the equation 
\[
  \frac{w^{1-\delta }  \Omega^\delta }{1-   \frac{ \kappa  }{   c^2 }  w^{1- \delta }\Omega^{1+\delta } +\frac{\Psi }{   c^2}  }  = \frac{V^2}{\left( \Omega -\frac{\omega }{c^2}\right) \left( 1-\frac{V^2}{c^2}\right) }.  
\]
    
    In  the Newtonian limit --- the zeroth order of the post-Newtonian expansion  (0PN)  --- one arrives at  $\Omega_0 = w/r^\frac{2}{1- \delta }$.
   
    The stability requirement of Ref. \refcite{Tassoul} imposes the condition  $  \delta \le  0$.    These  two constants, $w$ and $\delta $,  can be given apriori.    Let us remark at this point that the rotation law (\ref{momentum}), and consequently the Newtonian rotation   $\Omega_0 = w/r^\frac{2}{1- \delta }$, applies
primarily to single rotating toroids  and toroids rotating around black holes. In the case of rotating stars one would have to construct a special differentially rotating law, with the aim to avoid singularity at the rotation axis. 

The two limiting cases $\delta  =0$  and $ \delta =-\infty $ correspond to the constant angular momentum per unit mass  ($\Omega_0 =w/r^2$)  and the rigid rotation ($\Omega =w$), respectively. The Keplerian rotation is given by $ \delta =-1/3$ and $w^2=GM$, where $M$ is a mass (Ref. \refcite{MM}).

The   1PN approximation  corresponds to the choice of metric exponents
$\alpha =\beta =-\nu =-U$ with  $|U|\ll c^2$. Define $\omega \equiv r^{-2} A_\phi$; 
it appears that  $A_\phi $ satisfies  $
\Delta A_\phi -2\frac{\partial_rA_\phi }{r}= -16 \pi G r^2 \rho_0 \Omega_0 .$  

The spatial part of the metric 
\begin{eqnarray}
d s^2 &=&  -  \left(1+\frac{2U}{c^2} +\frac{2U^2}{c^4}\right) (d x^0)^2
-2  c^{-3} A_\phi d x^0 d \phi +\nonumber \\
&&\left(1-\frac{2U}{c^2}  \right)  \left( d r^2 + d z^2 + r^2 d \phi^2\right)    .
\label{metric1}
\end{eqnarray} 
is conformally flat.   
Notice that in  the Newtonian gauge imposed in  (\ref{metric1}) the geometric distance to the rotation axis   
is given by $\tilde r=r(1-U_0/c^2)+\mathcal{O}(c^{-4 })$.  The relevant post-Newtonian  approximation has been discussed  in Ref. \refcite{MM};
we  shall extract from therein   information on the general-relativistic corrections to
the angular velocity.   

The      angular velocity, up to the terms ${ \mathcal{O}(c^{-4})}$ is given by 
 \begin{eqnarray}
\label{angular velocityc2}
\Omega = \frac{w}{\tilde r^{2/(1-\delta)}}    -\frac{2}{c^2(1 -  \delta )} \Omega_0 \left( U_0+\Omega_0^2r^2\right)  &&\ + \frac{A_\phi}{r^2 c^2\left( 1- \delta \right)}   -  \frac{4}{c^2(1 -  \delta )} \Omega_0 h_0 .
\end{eqnarray}
This expression can be reduced to 
 $\Omega = \frac{w}{\tilde r^{2/(1-\delta)}}+\frac{A_\phi}{r^2 c^2\left( 1- \delta \right)}-  \frac{4}{rc^2(1 -  \delta )} \Omega_0 h_0$, in the case of test fluids, at the symmetry plane $z=0$. For the   massless dust, in the Schwarzschild geometry,  we get    $\Omega =\Omega_0  =\frac{w}{\tilde r^{3/2}}$.
 
 The first term in (\ref{angular velocityc2}) is simply the Newtonian rotation law rewritten as a function of the geometric distance from the rotation axis. 
 The second term  (denoted as $\Omega_\mathrm{1,NK}$ in Fig.1) vanishes at the plane of symmetry, $z=0$, for circular Keplerian motion of test fluids in the monopole  potential $-GM/\sqrt{r^2+z^2}$.  It is sensitive  both to the contribution of the disk self-gravity at the plane $z=0$ and the deviation  from the strictly Keplerian motion.
 The third  term ($\Omega_\mathrm{1,geo}$ in Figs 1 and 2) is responsible for the geometric frame dragging. The last    term ($\Omega_\mathrm{1,dyn}$ in Fig.\ 1) represents the recently discovered   anti-dragging effect;  it  agrees (for the monomial angular velocities $\Omega_0 =wr^{-2/(1- \delta )}$) --- with the result obtained earlier in Ref. \refcite{JMMP}.

  In the following discussion we assume $w>0$, which means $\Omega_0 >0$, but the reasoning is symmetric under the parity operation $w\rightarrow -w$.
The specific enthalpy  $h\ge 0$ is nonnegative, thence    $- \frac{4\Omega_0h_0}{1-\delta } $ is nonpositive --- the discovered in Ref.\ \refcite{JMMP}   instantaneous 1PN dynamic  reaction slows  the motion: it ``anti-draggs'' a system.
 In contrast to that, the well known geometric term with $A_\phi$ is positive \cite{JMMP}, and the contribution $\frac{A_\phi}{r^2\left( 1 - \delta \right)}$ to the angular velocity is positive --- it pushes a rotating fluid body forward.  Thus  the last two terms in  (\ref{angular velocityc2}) counteract.  
 
The  specific enthalpy $h_0$ vanishes for dust, hence dust test bodies   are  exposed only  to the frame dragging.  For  the rigid (uniform) rotation   the correction terms  are proportional to $1/(1-\delta )$ and they vanish, because     $\delta = - \infty$.   

Figure 1 displays contributions of each of the particular effects at the symmetry plane $z=0$ of the disk; notice that the anti-dragging can prevail the geometric dragging.
Figure 2 in turn demonstrates that the sum of all three contributing effects can be much larger than the geometric counterpart.

 \begin{figure}[h]
\begin{center}
\includegraphics[width=3.1in]{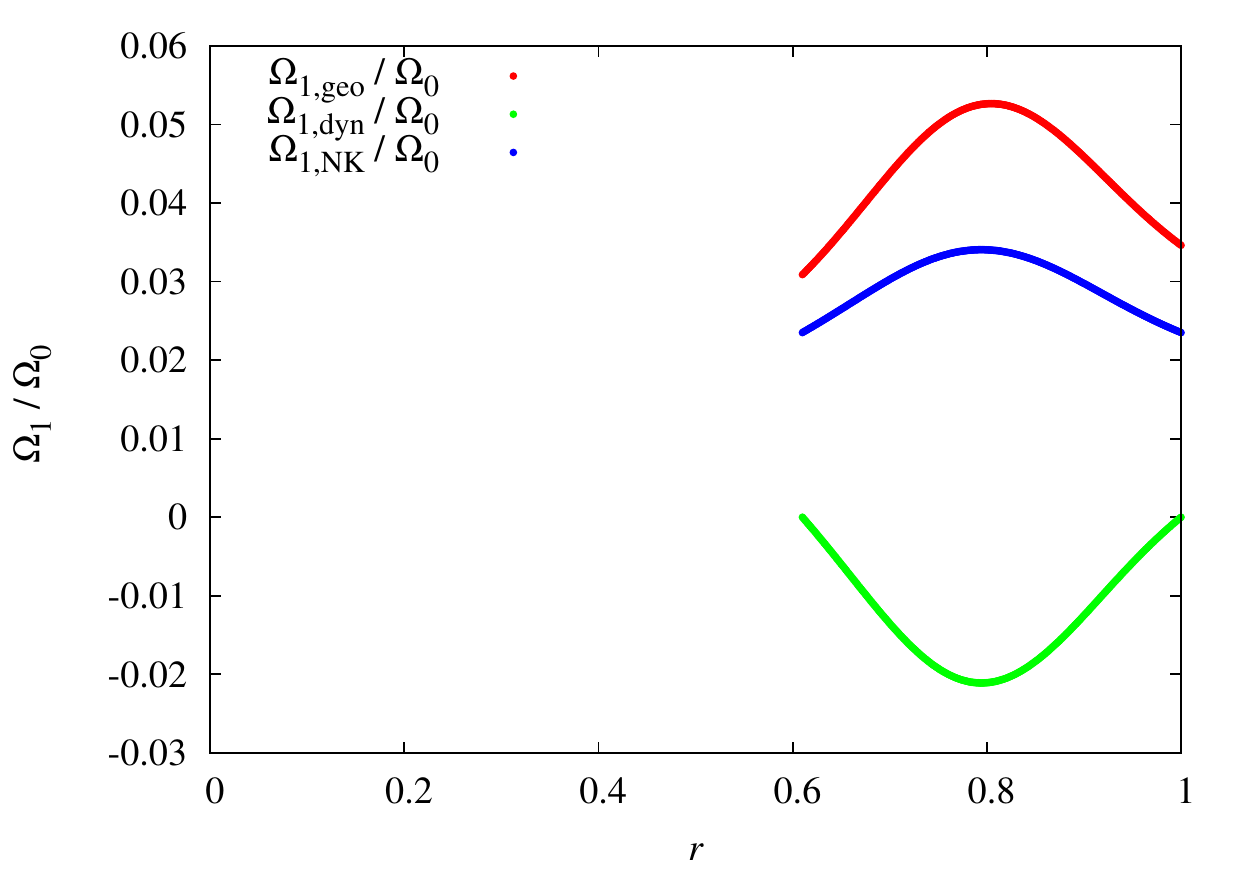}
\end{center}
\caption{The contributions of particular effects within the disk, $z=0$.}
\label{aba:fig1}
\end{figure}
 
 \begin{figure}[h]
\begin{center}
\includegraphics[width=3.1in]{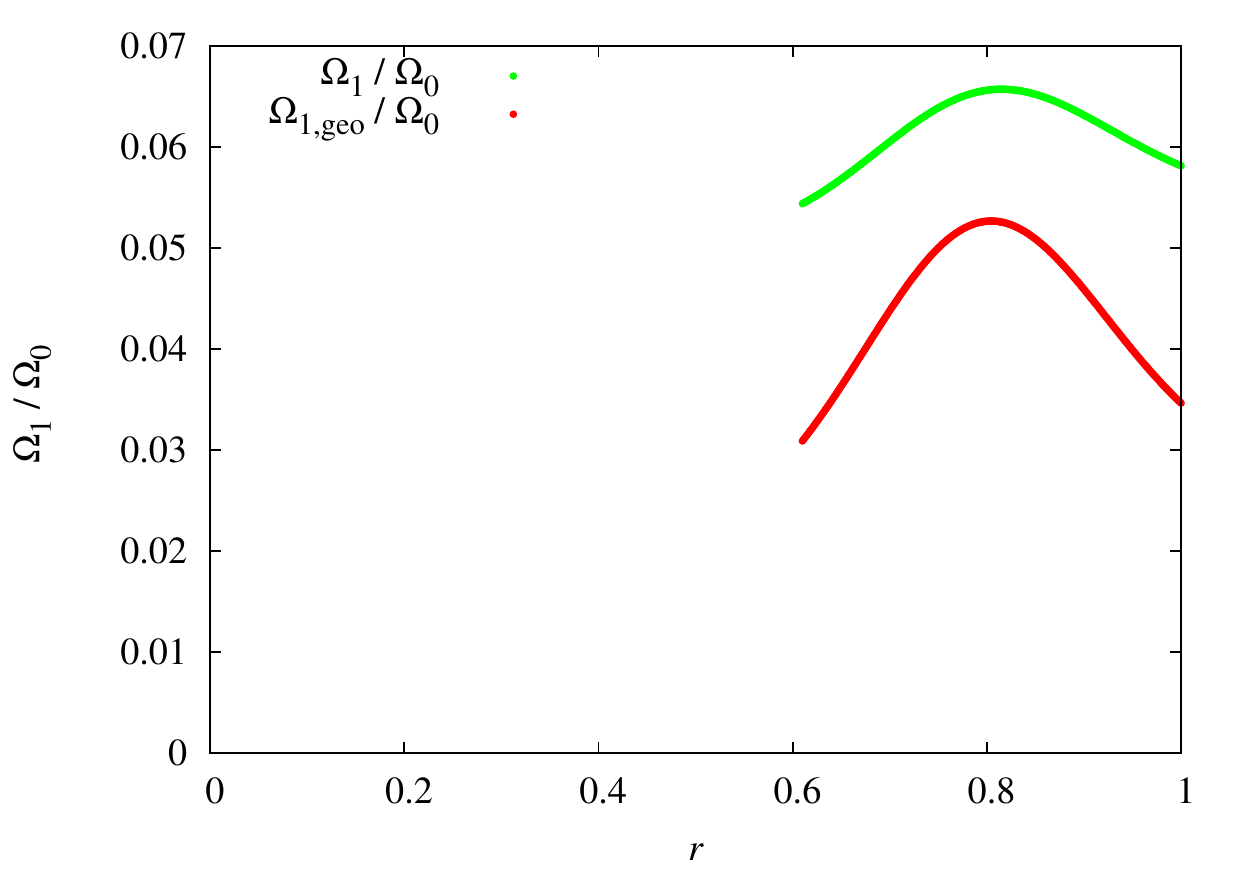}
\end{center}
\caption{The combined effect (the green line) and the geometric dragging (red line) within the dist,  $z=0$. Here $\Omega_1 = \Omega_\mathrm{1,NK}+\Omega_\mathrm{1,dyn}+\Omega_\mathrm{1,geo}$ }
\label{aba:fig2}
\end{figure}

\section{Concluding remarks}

We present  general-relativistic rotation laws and derive a full form of  the new weak-fields  effects, including  the recent dynamic anti-dragging effect of  Ref. \refcite{JMMP}.  The latter   can be robust according to the numerics of Ref.\  \refcite{JMMP}, but the ultimate conclusion requires a fully general-relativistic treatment with the new  rotation laws.    It is conjectured that in some active galactic nuclei \cite{Moran} the   frame dragging can manifest through  the   Bardeen-Petterson effect \cite{Bardeen-Petterson}. The two other effects may  lead    to its observable modifications    in  black hole systems with heavy  disks. 
In the weak field approximation    the angular velocity of toroids depends primarily on the distance from the rotation axis --- as in the Newtonian hydrodynamics --- but  the  weak fields contributions  can make the rotation curve dependent on the height above the symmetry plane of rotation.   
 
The new  rotation laws would allow   the investigation  of self-gravitating fluid bodies in the regime of strong gravity for general-relativistic versions of Newtonian rotation curves. 
In particular, they  can be used in order to describe stationary heavy  disks in tight accretion systems with central black holes, for instance in products  of the merger of compact binaries  (pairs of black holes and neutron stars) (Refs.\ \refcite{Pan_Ton_Rez} and \refcite{Lovelace}), but they might   exist also in some active galactic nuclei.  

These  rotation laws can be applied to the study of various open problems in   the post-Newtonian perturbation scheme of general-relativistic hydrodynamics --- the investigation of    convergence of the post-Newtonian pertubation scheme, as well  as the existence of  the Newtonian and post-Newtonian limits of   solutions. They can be used to test the accuracy of recent inequalities relating the angular momentum, the mass and area of black holes \cite{Dain,Khuri}.

\end{document}